\documentclass[a4paper,dvips,12pt]{article}
\usepackage{epsfig,graphics,graphicx}
\begin{document}

\pagestyle{myheadings}
\markright{\it CO02-6}
\vskip.5in
\begin{center}

\vskip.4in {\Large\bf Radiative corrections to the Casimir energy 
in the $\lambda|\phi|^{4}$ model under quasi-periodic boundary conditions}
\vskip.3in

F.\ A.\ Barone\footnote{Email: \tt fbarone@cbpf.br } \\
Centro Brasileiro de Pesquisas F\'{\i}sicas \\
Rua Dr.\ Xavier Sigaud, 150, Urca, 22290-180 Rio de Janeiro, RJ, Brazil
\\ 
R.\ M.\ Cavalcanti\footnote{Email: \tt rmoritz@if.ufrj.br }
and C.\ Farina\footnote{Email: \tt farina@if.ufrj.br }\\
Instituto de F\'{\i}sica, Universidade Federal do Rio de Janeiro \\
Caixa Postal 68528, 21941-972 Rio de Janeiro, RJ, Brazil
\end{center}
%

\vskip.2in
\begin{abstract}
We compute the first radiative correction to the 
Casimir energy in the $(d+1)$-dimensional 
$\lambda|\phi|^{4}$ model submitted to quasi-periodic 
boundary conditions in one spatial direction. Our results agree 
with the ones found in the literature for periodic and anti-periodic 
boundary conditions, special cases of the quasi-periodic boundary conditions.
\end{abstract}


The idea of introducing an arbitrary parameter in order to interpolate 
continuously distinct theories is not new in the literature. For instance, 
fermionic and bosonic partition functions can be obtained as particular 
cases of  more general ones which are computed assuming that 
these fields satisfy a more general boundary conditions (BC) in the 
imaginary time, where the fields acquire a phase $e^{i\theta}$ whenever $\tau\rightarrow\tau+\beta$ ($\beta=1/T$) \cite{Borges98-99} 
(for non-relativistic partition functions 
see Ref.\ \cite{FarinaHenrique}). Periodic and antiperiodic BC
(for bosons and fermions, respectively) correspond to $\theta=0$ and $\theta=\pi$.

It can be shown that the same effects 
can be obtained if instead of introducing the parameter $\theta$, we couple 
the charged field (bosonic or fermionic) appropriately with a constant 
gauge potential of the form $(A_0,{\bf 0})$, which cannot be gauged away 
due to the compactification in the $x_0$-direction, introduced to take 
into account the thermal effects.

Similarly, we can consider that the field under study is submitted 
to quasi-periodic BC in a space dimension, which interpolates
the periodic and antiperiodic ones. Analogously, the introduction of 
the interpolating parameter is equivalent to coupling the charged field 
with a constant gauge field with a non-vanishing component along the 
space-dimension which is assumed to be compactified \cite{Actor}. 

In this work we discuss the effects of an interpolating BC in the 
vacuum energy of a complex scalar field. More precisely, we compute the 
$O(\lambda)$ correction to the 
Casimir energy of a complex scalar field whose dynamics is
described by the (Euclidean) lagrangian density\footnote{Conventions:
$\hbar=c=1$; Greek indices vary from 0 to d; summation over repeated indices 
is understood unless explicitly stated.}
\label{Lagrangianatotal}
\begin{equation}
{\cal L}_{\rm E}=|\partial_{\mu}\phi|^{2}+m^{2}|\phi|^{2}+\lambda|\phi|^{4}
+{\cal L}_{\rm ct},
\end{equation}
where ${\cal L}_{\rm ct}$ contains the renormalization counterterms,
and subject to quasi-periodic boundary conditions in the $x_d$-direction, i.e.,
\begin{equation}
\label{condcontorno}
\phi(x_0,x_1,\ldots,x_d+a)=e^{i\theta}\phi(x_0,x_1,\ldots,x_d), 
\qquad 0\le\theta<2\pi.
\end{equation}
In previous works we performed similar calculations for Dirichlet-Dirichlet, 
Neumann-Neumann \cite{BCF1},
and Dirichlet-Neumann \cite{BCF2} boundary conditions.

The Casimir energy (per unit area) of a free field 
subject to those boundary conditions
was computed in \cite{Jayme1}
in the $(3+1)$-dimensional case; the result is
\begin{equation}
\label{E0}
{\cal E}_{\theta}^{(0)}(a)\Big|_{d=3}=-\frac{m^2}{\pi^2 a}
\sum_{n=1}^{\infty}\frac{1}{n^2}\,\cos(n\theta)\,K_2(nma),
\end{equation}
where $K_{\nu}(z)$ is the modified Bessel.
In the massless limit Eq.\ (\ref{E0}) becomes
\begin{equation}
{\cal E}_{\theta}^{(0)}(a)\Big|_{d=3,\,m=0}=-\frac{2}{\pi^2a^3}\sum_{n=1}^{\infty}
\frac{\cos(n\theta)}{n^4}
=\frac{2\pi^2}{3a^3}\,B_4\left(\frac{\theta}{2\pi}\right)\qquad(0\le\theta<2\pi),
\end{equation}
where $B_4(x)=x^4-2x^3+x^2-1/30$ is the Bernoulli polynomial
of fourth degree \cite{GR} (see Fig.\ 1).
\begin{figure}[hbt]
\begin{center}
\includegraphics*[scale=0.5, viewport=-150 -5 1000 500]{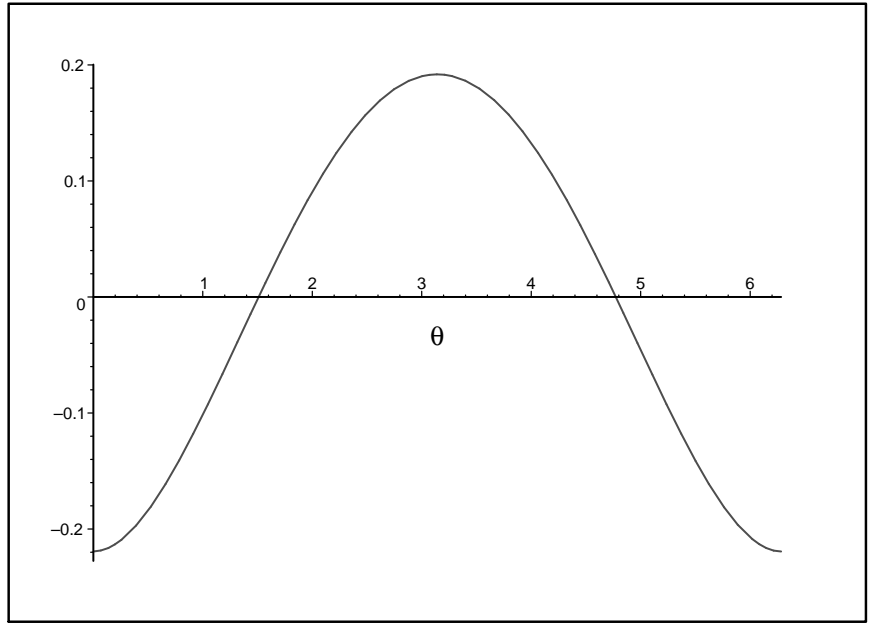}
\end{center}
\caption{$a^3{\cal E}_{\theta}^{(0)}(a)$ as a function of $\theta$
for a massless field in ($3+1$) dimensions.}%
\label{fig1}
\end{figure}

The $O(\lambda)$ correction to (\ref{E0}) is formally given by
\begin{equation}
\label{defcalE}
{\cal E}^{(1)}=\int_{0}^{a}dx_{d}\left[\lambda G^{2}(x,x)+\delta m^{2}\,G(x,x)+\delta\Lambda\right],
\end{equation}
where $G(x,x')$ is the Green's function of the free theory (i.e.,
with $\lambda=0$, but obeying the boundary conditions),
$\delta m^2$ is the radiatively induced shift in the mass parameter,
and $\delta\Lambda$ is the shift in the cosmological constant (i.e.,
the change in the vacuum energy which is due solely to the
interaction, and not to the confinement).

The spectral representation of $G(x,x')$ in $d+1$
dimensions is given by
\begin{equation}
G(x,x')=\frac{1}{a}\int{d^dk\over(2\pi)^{d}}\,e^{i{\bf k}\cdot({\bf x}-{\bf x}')} 
\sum_{n=-\infty}^{\infty}\frac{e^{iq_n(z-z')}}
{{\bf k}^{2}+q_{n}^{2}+m^2}\ ,
\end{equation}
where ${\bf x}=(x_0,\ldots,x_{d-1})$, $z=x_d$, and $q_n=(2n\pi+\theta)/a$.
$G(x,x')$ diverges when $x'\to x$ for $d\ge 1$, therefore a regularization
prescription is needed in order that $G(x,x)$ makes sense. We shall
compute it using dimensional regularization; the result is
\begin {equation}
\label {zxc1}
G(x,x)={\Gamma\left(1-d/2\right)\over(4\pi)^{d/2}a}
\sum_{n=-\infty}^{\infty}\omega_{n}^{d-2}\qquad(d<1),
\end {equation}
where $\omega_{n}=\sqrt{q_n^{2}+m^2}$. Inserting this result
into Eq.\ (\ref{defcalE}) and arranging terms, we obtain
\begin{equation}
\label{asd}
{\cal E}^{(1)}_{\theta}={\lambda\over a}\left[{\Gamma(1-d/2)\over(4\pi)^{d/2}}
\sum_{n=-\infty}^{\infty}\omega_{n}^{d-2}
+{a\,\delta m^{2}\over 2\lambda}\right]^{2}+a\left[\delta\Lambda 
-{(\delta m^{2})^{2}\over 4\lambda}\right].
\end{equation}

In order to compute the sum that appears in Eq.\ (\ref{asd})
it is convenient to reexpress it as
\begin{equation}
\label{sum}
\sum_{n=-\infty}^{\infty}\omega_{n}^{d-2}=\left({2\pi\over a}\right)^{d-2}
{\cal D}\left({2-d\over 2},{ma\over 2\pi},{\theta\over 2\pi}\right),
\end{equation}
where ${\cal D}$ is defined as
\begin{equation}
{\cal D}\left(s,\nu,\frac{\theta}{2\pi}\right):=
\sum_{n=-\infty}^{\infty}\left[\nu^2+\left(n+\frac{\theta}{2\pi}\right)^2
\right]^{-s},\qquad{\rm Re}(s)>1/2.
\end{equation}
The function ${\cal D}$ has an analytic continuation to the
whole complex $s$-plane given by \cite{FarinaHenrique}
\begin{equation}
\label{extanalitica}
{\cal D}\left(s,\nu,\frac{\theta}{2\pi}\right)=
\frac{\sqrt{\pi}\,\nu^{1-2s}}{\Gamma(s)}\left[\Gamma\left(s-\frac{1}{2}\right)
+4\sum_{n=1}^{\infty}\cos(n\theta)\,\frac{K_{1/2-s}(2n\pi\nu)}
{(n\pi\nu)^{1/2-s}}\right],
\end{equation}
with simple poles at $s=1/2,-1/2,-3/2,\ldots$ Inserting Eqs.\ (\ref{sum})
and (\ref{extanalitica}) into Eq.\ (\ref{asd}) yields

\begin{eqnarray}
\label{Ereg}
{\cal E}^{(1)}_{\theta}&=&{\lambda\over a}\left\{\frac{am^{d-1}}{(4\pi)^{(d+1)/2}}
\left[\Gamma\left(\frac{1-d}{2}\right)+4\sum_{n=1}^{\infty}\cos(n\theta)\,
\frac{K_{(d-1)/2}(nma)}{(nma/2)^{(d-1)/2}}\right]
+{a\,\delta m^{2}\over 2\lambda}\right\}^{2}
\nonumber \\
& &+a\left[\delta\Lambda 
-{(\delta m^{2})^{2}\over 4\lambda}\right].
\end{eqnarray}

Let us now fix the renormalization conditions for $\delta m^2$ and
$\delta\Lambda$. The former is fixed by imposing that the one-loop self-energy, 
given by $\Sigma^{(1)}(x,x)=2\lambda G(x,x)+\delta m^2$,
is finite and satisfies $\lim_{a\rightarrow\infty}\Sigma^{(1)}(x,x)=0$.
In addition, we require that $\delta m^2$ be independent of $a$.
All these conditions are fulfilled by taking
\begin{equation}
\label{dm2}
\delta m^2=-\lambda\,\frac{2m^{d-1}}{(4\pi)^{(d+1)/2}}\,
\Gamma\left(\frac{1-d}{2}\right).
\end{equation}
To fix $\delta\Lambda$ we also require that it does not depend on $a$,
and that the Casimir energy per unit volume, ${\cal E}/a$, vanishes
as $a\to\infty$. This is attained by taking 
\begin{equation}
\label{dL}
\delta\Lambda=\frac{(\delta m^2)^2}{4\lambda}\ .
\end{equation}
Inserting (\ref{dm2}) and (\ref{dL}) into Eq.\ (\ref{Ereg})
we finally arrive at the desired result:
\begin{equation}
\label{finald}
{\cal E}^{(1)}_{\theta}(a)=\frac{\lambda m^{d-1}}{2^{d-1}\pi^{d+1}a^{d-2}}
\left[\sum_{n=1}^{\infty}\cos(n\theta)
{K_{(d-1)/2}(man)\over n^{(d-1)/2}}\right]^{2}.
\end{equation}
In the special case of $d=3$ spatial dimensions it yields
\begin{equation}
\label{final3d}
{\cal E}^{(1)}_{\theta}(a)\Big|_{d=3}=\frac{\lambda m^2}{4\pi^4 a}
\left[\sum_{n=1}^{\infty}\cos(n\theta)\,\frac{K_1(nma)}{n}\right]^2.
\end{equation}
If we further take the limit $m\to 0$ we obtain
\begin{equation}
{\cal E}^{(1)}_{\theta}(a)\Big|_{d=3,\,m=0}=\frac{\lambda}{4\pi^4 a^3}
\left[\sum_{n=1}^{\infty}\frac{\cos(n\theta)}{n^2}\right]^2
=\frac{\lambda}{4a^3}\left[B_2\left(\frac{\theta}{2\pi}\right)\right]^2
\qquad(0\le\theta<2\pi),
\end{equation}
where $B_2(x)=x^2-x+1/6$ is the Bernoulli polynomial of second degree
\cite{GR} (see Fig.\ 2).
\begin{figure}[htb]
\begin{center}
\includegraphics*[scale=0.77, viewport=-100 -5 1000 300]{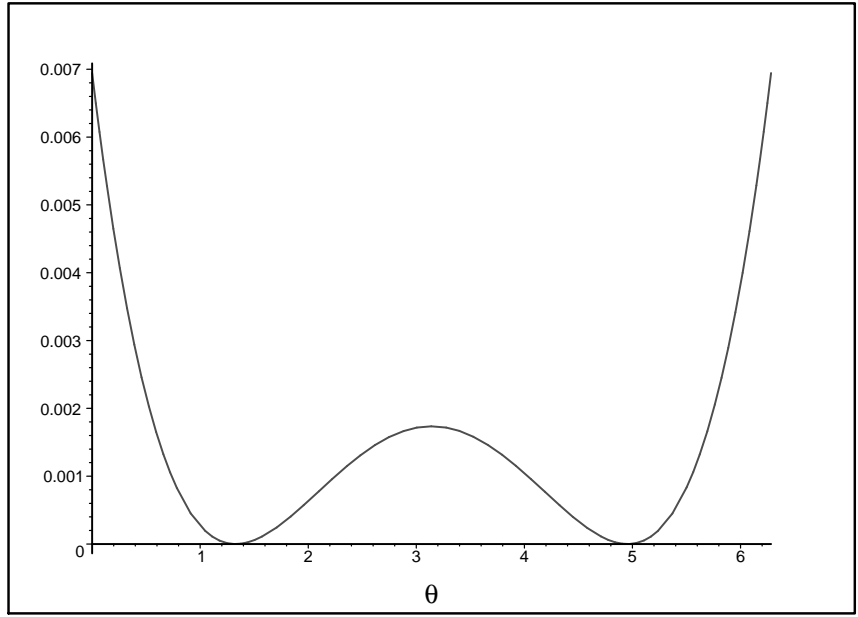}
\end{center}
\caption{$\lambda^{-1}a^3{\cal E}_{\theta}^{(1)}(a)$ as a function of $\theta$
for a massless field in ($3+1$) dimensions.}%
\label{fig1}
\end{figure}

By taking the limit $m\to 0$ in Eq.\ (\ref{finald}) we also 
recover results found in the literature 
for the periodic ($\theta=0$) and antiperiodic ($\theta=\pi$)
boundary conditions \cite{KrechDietrichPRA92}.


\section*{Acknowledgments}

F.A.B.\ is supported by FAPERJ. R.M.C.\ is supported by CNPq.
C.F.\ is partially supported by CNPq.



\end{document}